\documentclass[showpacs,amsfonts,amsmath,amssymb,aps,pra,reprint,groupedaddress]{revtex4-1}

\usepackage{units}
\usepackage{graphicx}
\usepackage{dcolumn}
\usepackage{bm}
\begin{document}


\title{Rapid Accurate Calculation of the $s$-wave Scattering Length}


\author{Vladimir V. Meshkov}
\author{Andrey V. Stolyarov}
\email[Email me at: ]{avstol@phys.chem.msu.ru}
\affiliation{Department of Chemistry, Moscow State University, Moscow, 119991, Russia}

\author{Robert J. \surname{Le Roy}}
\affiliation{Guelph-Waterloo Center for Graduate Work in Chemistry and
 Biochemistry, University of Waterloo, Waterloo, Ontario~N2L~3G1, Canada}


\date{\today}

\begin{abstract}


Transformation of the conventional radial Schr{\" o}dinger equation defined
on the interval $\,r\in[0,\infty)$ into an equivalent form defined on the
finite domain $\,y(r)\in [a,b]\,$ allows the $s$-wave scattering length
$a_s$ to be exactly expressed in terms of a logarithmic derivative of the
transformed wave function $\phi(y)$ at the outer boundary point $y=b$, which
corresponds to $r=\infty$.  In particular, for an arbitrary interaction
potential that dies off as fast as $1/r^n$ for $n\geq 4$, the modified
wave function $\phi(y)$ obtained by using the two-parameter mapping function $r(y;\bar{r},\beta) =
\bar{r}\left[1+\frac{1}{\beta}\tan(\pi y/2)\right]$ has no singularities, and
$$a_s=\bar{r}\left[1+\frac{2}{\pi\beta}\frac{1}{\phi(1)}\frac{d\phi(1)}{dy}
\right]~.$$
For a well bound potential with equilibrium distance $r_e$, the optimal mapping parameters
are $\,\bar{r}\approx r_e\,$ and $\,\beta\approx \frac{n}{2}-1$.
An outward integration procedure based on Johnson's log-derivative
algorithm [B.R.\ Johnson, J.\ Comp.\ Phys., \textbf{ 13}, 445 (1973)]
combined with a Richardson extrapolation procedure is shown to readily
yield high precision $a_s$-values both for model Lennard-Jones ($2n,n$)
potentials and for realistic published potentials for the Xe--e$^-$,
Cs$_2(a\,^3\Sigma_u^+$) and $^{3,4}$He$_2(X\,^1\Sigma_g^+)$ systems.
Use of this same transformed Schr{\"o}dinger equation was previously shown
[V.V. Meshkov {\em et al.} Phys.\ Rev.\ A, {\bf 78}, 052510 (2008)] to ensure
the efficient calculation of all bound levels supported by a potential,
including those lying extremely close to dissociation.


\end{abstract}


\maketitle



\section{Introduction}

The so-called the $s$-wave scattering length $a_s$ is a key parameter for
describing the interaction of particles at very low collision energies.
In particular, the two-body collision problem is completely specified
by the scattering length in the low temperature limit where the elastic
cross section becomes $\sigma_e = 4\pi a_s^2$ \cite{landau, mott}.
Many of the properties of a Bose-Einstein condensate \cite{bloch08}
also depend only on the scattering length.  In particular, the chemical
potential of a uniform Bose gas is simply proportional to the $a_s$-value
\cite{hutson06}, namely: $\mu_{Bose} = a_s n 4\pi{\hbar}^2/m$, where $n$
is the number density and $m$ is the atomic mass \cite{pethick02}.  Thus,
positive $a_s$ values correspond to overall repulsive interactions while
negative values correspond to attractive ones.

The scattering length is asymptotically related to the scattering phase
shift $\eta_s(k)$ by the expression \cite{landau, mott, rodberg67}:
\begin{equation}\label{eq:phaseshift}
a_s=-\lim_{k\to 0}\frac{\tan \eta_s(k)}{k}
\end{equation}
in which $k$ is the relative wave vector of the colliding particles.  The
scattering length can be also determined from the wavefunction $\psi(r)$
of the radial Schr\"{o}dinger equation
\begin{eqnarray}  \label{eq:schrod0}
  -\,\frac{d^2\psi(r)}{dr^2} ~&=&~ Q(r)\psi(r)   \\
  Q(r) ~&\equiv&\,  -\,\frac{2\mu}{{\hslash}^2}\, U(r)   \nonumber
\end{eqnarray}
solved at zero energy $E=0$ with the inner boundary condition $\psi(0)=0$.
It has been proved \cite{landau} that for arbitrary potentials $U(r)$
that obey the asymptotic condition:
\begin{eqnarray}  \label{eq:asympt1}
\lim_{r\to \infty}r^nU(r)~=~0 \quad{\rm for} \quad n>3 ~~,
\end{eqnarray}
the wavefunction $\psi(r)$ has the linear asymptotic form
\begin{eqnarray}  \label{eq:asympt0}
\psi(r) ~\simeq~ S\,(r-a_s)~~, \quad r\to \infty ~~,
\end{eqnarray}
in which the slope $\,S\,$ is a constant.  It can be seen from
Eq.(\ref{eq:asympt0}) that the scattering length physically corresponds to
the distance where the continuation of this asymptotic straight line crosses
the $r$-axis.  It is clear that the value of $a_s$ will be strongly depend
on the interaction potential $U(r)$ and on the reduced mass of the colliding
particles $\mu$.  It is also well known that the scattering length $a_s$
is approximately related to the binding energy $E^b_{v_\textrm {max}} \geq
0\,$ of the highest bound level ($v=v_{\rm max}$)
of the interaction potential $U(r)$ \cite{roman65}:
\begin{eqnarray}\label{eq:Energylast}
a_s~\approx~ \sqrt{\frac{\hslash^2}{2\mu\,E^b_{v_\textrm {max}}}}
\end{eqnarray}
It is clear that $a_s\to +\infty$ as $\,E^b_{v_\textrm {max}}\to 0$,
while $a_s$ will take on large negative values if the potential well
is almost deep enough to support one more bound level.  In other words,
depending on the potential and the reduced mass, the scattering length
can have any value on the interval $(-\infty,+\infty)$.

Accurate determination of scattering lengths determined from photoassociation
of ultracold colliding atoms \cite{jones06} offers the possibility of
constructing reliable empirical interatomic potentials all the way to the
dissociation limit \cite{Abraham95, Julienne1995}.  However, rapid and robust
methods for performing scattering length calculations are required to make
such an inversion procedure feasible.  Furthermore, a high degree of accuracy
in the calculated $a_s$-values is required in order to provide the accurate
derivatives of the scattering length with respect to the parameters defining
the potential that are required for performing such fits \cite{gutierrez84}.

There are a number of existing schemes for calculating scattering lengths.
However, calculation of the scattering phase shift $\eta_s(k)$ as a
function of energy \cite{landau, mott} in order to apply the low-energy
extrapolation of Eq.\,(\ref{eq:phaseshift}) leads to large uncertainties,
especially for large $a_s$ values, since $\eta_s(k)$ becomes a very
steep function in the vicinity of $k=0$.  Moreover, the direct numerical
integration of the radial equation (\ref{eq:schrod0}) cannot be performed
easily because the asymptotic linear form of Eq.\,(\ref{eq:asympt0})
is only reached at very large distances $r$, typically thousands of \AA.
More robust methods \cite{gutierrez84, marinescu94, szmytkowski95} are
based on integration of Eq.\,(\ref{eq:schrod0}) to some finite matching
distance $r_m$, and then matching the resulting numerical wave function
$\psi(r_m)$ with an asymptotic counterpart for the long-range region which
is known analytically for particular kinds of potentials.  Alternatively,
the influence of a long-range interaction on the truncated apparent
$a_s$-value at some distance $r=r_m$ can be effectively corrected in the
framework of a secular perturbation theory expansion \cite{ouerdane03}.
These asymptotic methods often yield fairly accurate results, but not for
large $a_s$ values.  Moreover, they all require additional computational
effort to ensure the convergence of results with respect to an optimal $r_m$
value whose position is not well-defined \emph{a priori}.  These asymptotic
methods also cannot readily be used for potentials for which the asymptotic
solution is not available in closed form.  We also note that a very elegant
analytical formula for $a_s$ has been obtained using the semiclassical
approximation for the wavefunction $\psi(r)$ \cite{gribakin93}.  However,
while that formula can be useful for estimation purposes, its accuracy is
limited by its dependence on the WKB approximation \cite{landau, child},
so it does not suffice for many applications.

The key to the present method is the analytically exact transformation 
\cite{LiouvilleGreen, meshkov08} of the initial Schr{\"o}dinger equation (\ref{eq:schrod0}) defined on infinite
domain $r\in [0,\infty)$ into a modified radial equation defined by a reduced
variable $y(r)$ on a finite interval $y(r)\in [a,b]$ \cite{Ogilvie, Surkus}.
The resulting transformed equation can be solved very efficiently by
standard numerical methods \cite{Numerov, pseudospectral}. Moreover, we show that a special choice of
the mapping function $r(y)$ leads to a transformed wave function $\phi(b)$
that has no singularities on the interval $[a,b]$.  In this case the
scattering length can be expressed analytically in terms of the logarithmic
derivative of the solution at the (finite) end point $b$ and the associated
mapping parameters.  One efficient way to obtain the required logarithmic
derivatives of the wave function at $\,y=b$ is to numerically integrate the
resulting modified Ricatti equation using Johnson's log-derivative method
\cite{johnson73, johnson77} and apply a Richardson extrapolation to the
values obtained at the limit \cite{Richardson, NR}.  A FORTRAN code
applying this approach that has been applied both to model Lennard-Jones
($2n,n$) potentials and to realistic potentials for Cs$_2(a\,^3\Sigma_u^+)$ 
\cite{gribakin93, marinescu94, szmytkowski95}, Xe--e$^-$ \cite{czuchaj87} 
and $^{3,4}$He$_2(X\,^1\Sigma_g^+)$ \cite{aziz79} taken from
the literature has been provided as Supplementary Documentation \cite{EPAPS}.
An alternate version of the present approach based on the conventional Numerov
propagator \cite{Numerov} has been implemented in a `finite domain' version
of the general purpose bound-state/Franck-Condon code LEVEL \cite{LEVELas}.

\section{Adaptive Mapping Procedure for Scattering Length Calculations}
\subsection{Reduced variable transformation of the radial equation}

We begin by introducing a mapping function $y=y(r)$ that is a smooth,
monotonically increasing function of the radial coordinate $r$ and maps
$[0,\infty)$ onto the finite domain $[a,b]$.  The well-known substitution
\cite{meshkov08, LiouvilleGreen}
\begin{equation}  \label{eq:gy}
\psi(r(y))~=~\sqrt{g(y)}~\phi(y)~~,~~ \quad g(y)~\equiv~ \frac{dr}{dy}~>~0
\end{equation}
then transforms the conventional radial Schr{\"o}dinger equation
(\ref{eq:schrod0}) into the equivalent form
\begin{equation}  \label{eq:schrod}
\frac{d^2\phi(y)}{dy^2}~=~-\,\widetilde{Q}(y)~\phi(y)
\end{equation}
in which
\begin{equation} \label{eq:tildaO}
\widetilde{Q}(y)~\equiv~ g^2(y)~Q(r(y))~+~ F(y)
\end{equation}
whose additive term is defined as
\begin{equation}\label{eq:Fy2}
F(y)~\equiv~ \frac{~g^{\prime\prime}}{2g}~ -~\frac{3}{4} \left( \frac{g^{\prime}}{g}\right)^2
\end{equation}
Hereafter, a prime ($^{\prime}$) symbol denotes differentiation with
respect to our new radial variable $y$.

It should be noted that:
\begin{enumerate}
\item the modified wave function $\phi(y)$ is now defined on the finite
domain $[a,b]$;

\item the modified radial equation (\ref{eq:schrod}) is completely
equivalent to the initial one (\ref{eq:schrod0}), as long as $r(y)\in
(\mathcal{C}^3[a,b])$ is a monotonically increasing function;

\item a knowledge of the inverse analytical function $r(y)$ is
all that is required to accomplish the exact transformations of
Eqs.\,(\ref{eq:gy})-(\ref{eq:Fy2});

\item for a fixed mesh of equally spaced $y$ points on the interval $y\in
[a,b]$, the function $\rho(r) = dy/dr = 1/g(y)$ defines the density
distribution of mesh points of the initial coordinate $r$.

\end{enumerate}

It is clear that conventional finite-difference \cite{johnson73,
johnson77, NR, Numerov} and pseudospectral \cite{pseudospectral} methods can
be used straightforwardly for integrating the transformed radial equation
(\ref{eq:schrod}) out to the point $y=b$, as long as the modified wave
function $\phi(y)$ is not singular anywhere on the interval $[a,b]$, and
especially not at its end points.  Furthermore, it will be proved in the
next section that the scattering length $a_s$ is an explicit function of
logarithmic derivative of the wave function $\phi(y)$
\begin{equation} \label{eq:logderiv}
\xi(y)~\equiv~ \frac{\phi^{\prime}(y)}{\phi(y)}
\end{equation}
at the boundary $y=b$.  It is well-known that the transformation
(\ref{eq:logderiv}) converts the radial equation (\ref{eq:schrod}) into
the modified Ricatti equation
\begin{equation}\label{eq:riccati}
 \xi^{\prime}(y)~+~\widetilde{Q}(y)~+~\xi^2(y)~=~0
\end{equation}
which can be numerically integrated, for instance, using Johnson's efficient
log-derivative method \cite{johnson73, johnson77}.  More details regarding this
method are given in the Appendix.

\subsection{Developing a Formula for the Scattering Length}\label{formuladev}

Since Eq.\,(\ref{eq:asympt0}) shows that the scattering length can be
expressed formally in terms of the asymptotic behavior of the ordinary
radial wavefunction $\psi(r)$, we can write
\begin{eqnarray}\label{eq:asymptr}
a_s~ \simeq~ r~-~\frac{\psi(r)}{d\psi(r)/dr}~~, ~~\quad r\to \infty~~.
\end{eqnarray}
It therefore seems desirable to investigate the asymptotic behavior of the
modified function $\phi(y)$ near the outer boundary point $y=b$ corresponding
to $r\to \infty$, where
\begin{eqnarray}  \label{eq:modphi}
\phi(y)~ \simeq~ S~\frac{r(y)-a_s}{\sqrt{dr(y)/dy}}~~,\quad y\to b ~~.
\end{eqnarray}

Since we are mapping an infinite interval onto a finite one, it is clear
that the mapping function $r(y)$ must have a singularity at the upper end
of the interval.  To proceed, we assume that this singularity has the form
\begin{eqnarray}\label{eq:rasympt}
r(y)~\sim~ \frac{1}{(b-y)^\gamma}~~,~~ \quad y\to b ~~,
\end{eqnarray}
where necessarily $\gamma>0$, because the $r(y)$ function must go to
infinity as $y$ approaches $b$ in order to convert the finite domain $y\in
[a,b]$ into the infinite one $r\in [r_{\rm min},+\infty]$.

Inserting Eq.\,(\ref{eq:rasympt}) into Eq.\,(\ref{eq:modphi}) yields
\begin{eqnarray} \label{eq:leadterm}
\phi(y)~\simeq~ S~(b-y)^{\frac{1-\gamma}{2}} ~~,~~ \quad y\to b
\end{eqnarray}
and differentiating that result with respect to $y$ yields
\begin{eqnarray} \label{eq:leadterm1}
\phi^{\prime}(y)~\simeq~ S~\frac{1-\gamma}{(b-y)^{\frac{1+\gamma}{2}}}
  ~~,~~ \quad y\to b ~~.
\end{eqnarray}
From Eqs.(\ref{eq:leadterm}) and (\ref{eq:leadterm1}) it is immediately
clear that for both the modified wave function $\phi(y)$ and its derivative
$\phi^{\prime}(y)$ to be non-singular at the point $y=b$ (or $r=+\infty$),
necessarily $\gamma=1$.

Avoidance of this singularity at $y=b$ is clearly desirable if we are to
integrate the modified radial equation (\ref{eq:schrod}) accurately on the
whole interval $[a,b]$, so it is necessary to choose a mapping function
$r(y)$ that has only a simple pole at the end point $y=b$.  Let us assume
that $r(y)$ can be expressed as a Laurent series expansion about this point:
\begin{eqnarray} \label{eq:ry_to_b}
r(y)~=~ \frac{c_{-1}}{b-y}~+~c_0~+~ \sum_{i=1}^{\infty}c_i(b-y)^i
\end{eqnarray}
where $\,c_{-1}>0$. Inserting Eq.\,(\ref{eq:ry_to_b}) into
Eq.\,(\ref{eq:modphi}) and differentiating the resulting expression for
$\phi(y)$ with respect to $y$ yields, at the point $y=b$,
\begin{eqnarray} \label{eq:phi}
\phi(b)~=~ S\,\sqrt{c_{-1}}~~, \hspace{10mm}
\phi^{\prime}(b)~=~ \frac{S(a_s-c_0)}{\sqrt{c_{-1}}}~~,
\end{eqnarray}
which shows that
\begin{eqnarray} \label{eq:scatlen}
a_s~=~ c_{-1}\,\xi(b)~+~ c_0
\end{eqnarray}
is the desired relationship between the scattering length $a_s$ and
the log-derivative function at the outer end of the interval, $\xi(b)$.

To facilitate accurate calculation of the the required $\xi(b)$ value, it
is desirable that the function $\widetilde Q(y)$ of Eq.\,(\ref{eq:tildaO})
be nonsingular at the end point $y=b$.  It is easy to verify that the
$F(y)$ contribution is nonsingular at $y=b$, since
\begin{eqnarray} \label{eq:F(b)}
F(b)~=~-\,\frac{3c_1}{c_{-1}}
\end{eqnarray}
Moreover, from Eq.\,(\ref{eq:ry_to_b}) it is clear that in the limit
$\,y\to b\,$ (or $\,r\to \infty$),
\begin{eqnarray} \label{eq:y_to_b}
g(y)~\equiv~\frac{dr(y)}{dy}~\simeq~ \frac{c_{-1}}{(b-y)^2} ~\propto~ r^2 ~~.
\end{eqnarray}
This means that for any potential which dies off more slowly than $1/r^4$, the
product $g^2\,U(r)$ that comprises the main part of the function $\widetilde
Q(y)$ (see Eq.\,(\ref{eq:tildaO})) diverges in the limit $y\to b$ $(r\to
+\infty)$. In particular, if $\,U(r\to \infty) \simeq -\,C_n/r^n\,$, then
as $\,y\to b\,$,
\begin{eqnarray} \label{eq:g2U}
g^2\,U(y) ~&\simeq&~ \frac{2\mu}{\hslash^2}~\frac{C_n (b-y)^{n-4}}
{(c_{-1})^{n-2}}~~,
\end{eqnarray}
and hence
\begin{eqnarray}\label{eq:Qlim}
\widetilde Q(b) =~ \left\{ \begin{array}{c c}  F(b)~~~~  &  n > 4 \\[1ex]
\displaystyle \frac{2\mu}{\hslash^2}~\frac{C_n}{(c_{-1})^2} + F(b)~~ &  n=4
\\[2ex] \infty~~~~  &  n < 4  \end{array}  \right.   \nonumber
\end{eqnarray}
This means that, to calculate the modified wavefunction $\phi(y)$ and/or
its derivative $\phi^{\prime}(y)$ at the end of the range ($y=b$) for an
$n=4$ case requires explicit knowledge of the leading long-range induction
coefficient $C_4$, while for the more common $n=5$ or 6 cases it requires
only a knowledge of the mapping function $r(y)$ that defined $F(b)$.
In contrast, for $\,n < 4\,$ no satisfactory determination of $\phi(b)$
and/or $\phi'(b)$ can be achieved using the present approach.  This latter
result is consonant with the fact that the scattering length is not defined
for potentials that die off more slowly than $1/r^4$ \cite{landau}.

\subsection{Introduction of Mapping Functions}

To make practical use of our relationship (\ref{eq:scatlen}) between the
scattering length and the log-derivative function, it is necessary to
introduce an analytical mapping function with the Laurent expansion form
of Eq.\,(\ref{eq:ry_to_b}).  The simplest way to do this is to define the
mapping function as the first two terms of the Laurent series:
%
%
\begin{equation} \label{eq:simp_map}
r(y) = \frac{c_{-1}}{b-y} + c_0 ~~,
\end{equation}
since only the $c_{-1}$ and $c_0$ coefficients are required in
Eq.\,(\ref{eq:scatlen}).  In particular, setting $b=1$ and $\,c_{-1}/2= -
c_0 \equiv \bar{r},$ in (\ref{eq:simp_map}) yields a version of (omitting a
factor of 2) of the well-known Ogilvie-Tipping (\texttt{OT}) potential energy
expansion variable \cite{Ogilvie}:
\begin{equation} \label{eq:ogilvie}
y_{\texttt{OT}}(r;\bar{r}) = \frac{r-\bar{r}}{r+\bar{r}} ~~;\quad
y_{\texttt{OT}}\in[-1,1] ~~.
\end{equation}
For this case Eq.\,(\ref{eq:scatlen}) shows that the scattering length is
explicitly defined as
\begin{equation} \label{eq:scatogilvie_as}
a_s=\bar{r}\left[2\,\xi(1)-1\right]
\end{equation}
while the reciprocal mapping function is
\begin{equation} \label{eq:rscatogilvie}
r_{\texttt{OT}}(y)=\bar{r}\left(\frac{1+y}{1-y}\right )
\end{equation}
and
\begin{equation} \label{eq:gscatogilvie}
g_{\texttt{OT}}(y) = \frac{2\bar{r}}{(1-y)^2}= \frac{(r+\bar{r})^2}{2\bar{r}}
\end{equation}
Moreover, for this mapping function the additive term $F_{\texttt{OT}}(y) =
0$.

Use of the simple one-parameter mapping function of Eq.\,(\ref{eq:ogilvie})
does provide a reliable method for calculating $a_s$.  However, our
recent experience with applying this type of transformation to bound-state
problems \cite{meshkov08} suggests that better efficiency may be attained
using more flexible two-parameter functions.  One such function which was
successfully used for bound-state calculations is the two-parameters
($\bar{r}$, $\alpha>0$) {\v S}urkus variable \cite{Surkus}:
\begin{equation} \label{eq:surkus1}
y_{\rm S}(r;\bar{r},\alpha)~=~\frac{r^{\alpha}-\bar{r}^{\alpha}}{r^{\alpha}+
\bar{r}^{\alpha}}~~;\quad y_{S}\in[-1,1]
\end{equation}
for which the reciprocal mapping function is
\begin{eqnarray}\label{eq:surkus2}
r_{\rm S}(y)~=~\bar{r}\left(\frac{1+y}{1-y} \right)^{1/\alpha}
\end{eqnarray}
However, except for the special case $\alpha=1$ in which $y_{\rm
S}(r;\bar{r},\alpha)$ reduces to $y_{\texttt{OT}}(r;\bar{r})$, this mapping
function is not appropriate for scattering-length calculations, since when
$\alpha\neq 1$ the mapping function $r_{\rm S}(y)$ does not take on the
required Laurent series expansion form (\ref{eq:ry_to_b}) as $y\to 1$, so
the associated log-derivative wavefunction $\xi(y)$ becomes singular there.

In contrast, the two-parameters tangential function \cite{pseudospectral}
\begin{eqnarray}\label{eq:rtan}
r_{\rm tg}(y)~=~\bar{r}\left [1+\frac{1}{\beta}\tan\left(\frac{\pi y}{2}\right)
\right ]
\end{eqnarray}
defined on the interval $y \in  [(2/\pi)\tan^{-1}(-\beta),1]$, whose inverse is
the inverse tangent function
\begin{eqnarray} \label{eq:rarctan}
y_{\rm tg}(r;\bar{r},\beta)~=~\frac{2}{\pi}~\tan^{-1}\left[\beta\left(r/\bar{r}-1
\right)\right] ~~,
\end{eqnarray}
completely conforms to the required Laurent expansion form of
Eg.\,(\ref{eq:ry_to_b}) as $y\to 1$, since
\begin{eqnarray}\label{eq:Laurent}
r_{\rm tg}(y)~\simeq~ \frac{2\bar{r}}{\pi\beta(1-y)}+\bar{r}-
\frac{\pi\bar{r}}{6\beta}(1-y)+\dots
\end{eqnarray}
For this case
\begin{eqnarray} \label{eq:scatarctan}
a_s=\bar{r}\left[\frac{2\xi(1)}{\pi\beta}+1\right] ~~,
\end{eqnarray}
and it is easy to show that the additive contribution to Eq.\,(\ref{eq:tildaO})
is a constant,
\begin{eqnarray} \label{eq:arctanaddterm}
 F_{\rm tg}(y)~=~\frac{\pi^2}{4} ~~,
\end{eqnarray}
and that
\begin{eqnarray} \label{eq:arctag}
  g_{\rm tg}(y)&=&\frac{\pi \bar{r}}{2\beta}~\cos^{-2}\left(\pi y/2\right)  \\
           &=&\frac{\pi \bar{r}}{2\beta}\left [1+\beta^2\left(r/\bar{r}-1
               \right)^2\right ]~~. \nonumber
\end{eqnarray}

\subsection{Determination of Optimal Mapping Parameters}

The optimal values of the parameters defining the mapping function of
Eq.\,(\ref{eq:rtan}) may be expected to depend both
on the nature of the interaction potential $U(r)$ and on the particular
numerical method used for integration of the modified radial equation
(\ref{eq:schrod}).  When using Johnson's log-derivative method of integration
\cite{johnson73, johnson77}, the optimal mapping parameters $\bar{r}$ and
$\beta$ could be determined by minimizing the truncation error $\Delta \xi$
estimated using a Richardson extrapolation (RE) (\ref{eq:Richardson2}), or by
minimizing the difference between $\xi_{h_1}(y=b)$ and $\xi_{h_2}(y=b)$
values corresponding to two different integration step sizes $h_1$ and $h_2$,
\begin{equation} \label{eq:minRichardson}
\min |\xi_{h_1}-\xi_{h_2}| ~~.
\end{equation}
Note that it is tacitly assumed that the optimal parameter values do
not depend on the integration step size.  This has been confirmed by the
illustrative calculations presented in Section \ref{numtest}.  It will also
be shown there that the minimum of the functional (\ref{eq:minRichardson}) is
a rather smooth function of both mapping parameters, $\bar{r}$ and $\beta$.

\begin{figure}[t!]
\includegraphics{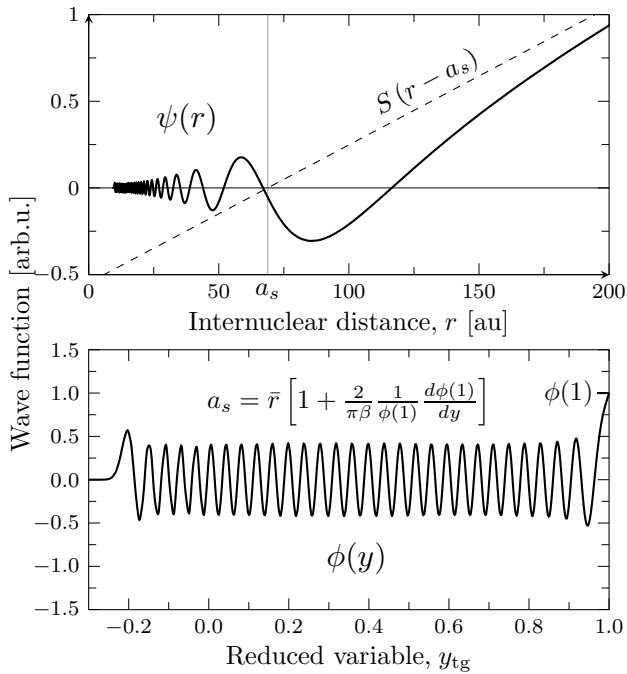}  \vspace{-2mm}
\caption{Comparison of the original $\psi(r)$ and modified
$\phi(y)=\psi/\sqrt{g_{\rm tg}}$ unbound wave functions at zero energy
for the $a\,^3\Sigma_u^+$ state of Cs$_2$ \cite{gribakin93}, as calculated
using the Numerov method \cite{Numerov, EPAPS}.  The scaling function
$g_{\rm tg}(r)$ was calculated from Eq.\,(\ref{eq:arctag})
using the mapping parameters $\bar{r}=r_e=12.0~[au]$ and
$\beta=2$. \label{fig:Cs2wavefunction}}
\end{figure}

It seems reasonable to assume that the optimal mapping function would
correspond to the situation in which
\begin{equation} \label{eq:optmapping}
\widetilde{Q}(y)~\equiv~ g^2(y)~Q(r(y))~+~ F(y) ~\approx~ const
\end{equation}
since it is well-known that minimal truncation errors arise in
particle-in-a-box problems where the potential $U(r)$ does not depend on $r$.
Furthermore, within the classically allowed region where $Q\gg 0$, the
additive term $F(y)$ can be approximately neglected \cite{landau, child}.
In this case Eq.\,(\ref{eq:optmapping}) reduces to $g^2\,Q\approx const$,
and inverting this condition (recalling that $g(y)\equiv dr/dy$) yields
\begin{eqnarray}\label{eq:WKB}
y_{\rm opt}(r>r_0)~\sim\int_{r_0}^r\sqrt{Q(r)}dr
\end{eqnarray}
Here $r_0$ is the left-hand (inner) classical turning point, which is the
root of the equation $U(r_0)=0$.

\begin{figure}[t!]
\includegraphics{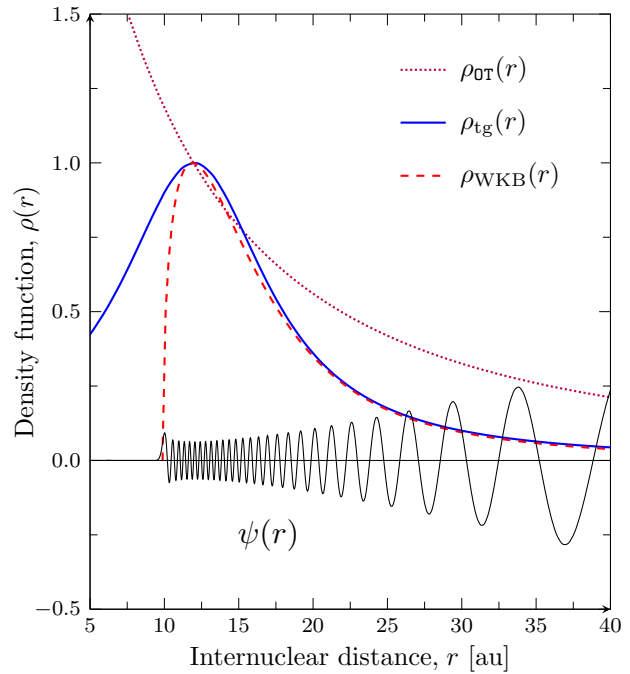}   \vspace{-2mm}
\caption{(Color online) Mapping density functions $\rho(r)=dy/dr=1/g(r)$
for the $a\,^3\Sigma_u^+$ state of Cs$_2$ \cite{gribakin93} calculated:
($i$) within the framework of the conventional WKB approximation
$\rho_{\rm WKB}=\sqrt{-U(r)}$ (\ref{eq:wkbopt}),~ ($ii$) using the one-parameter 
$y_{\texttt{OT}}(r;\bar{r})$ mapping function (\ref{eq:ogilvie}) with $\bar{r}=r_e=12.0~[au]$, and~ ($iii$) using
the two-parameters $y_{\rm tg}(r;\bar{r},\beta)$ mapping function (\ref{eq:rarctan}) 
with $\bar{r}=r_e$ and $\beta=2$.  $\psi(r)$ is the associated
zero-energy wavefunction. }\label{fig:Cs2dens} \end{figure}

Differentiating Eq.\,(\ref{eq:WKB}) with respect to $r$ yields
\begin{equation}\label{eq:wkbopt}
\rho_{\rm opt}~\equiv~\frac{dy_{\rm opt}}{dr}~ \sim~ \sqrt{Q(r)}~\equiv~
\sqrt{-\,\frac{2\mu}{\hslash^2}\, U(r)}
\end{equation}
For optimal mapping, therefore, the density function $\rho_{\rm opt}(r)$
should be close to $\sqrt{Q(r)}$ in the $Q(r)\gg 0$ region where the
original wavefunction $\psi(r)$ has its maximum oscillation frequency.
From Eq.\,(\ref{eq:wkbopt}) it follows that the optimal mapping does not
depend on the reduced mass $\mu$, because it appear in $Q(r)$ simply as a
multiplicative factor.  It is also easy to see that the mapping function
(\ref{eq:WKB}) transforms the modified wavefunction $\phi(y)$ into
the familiar particle-in-a-box form
\begin{eqnarray}\label{eq:sincos}
 \phi_{\rm opt}(y)\sim A\sin(ky)+B\cos(ky)
\end{eqnarray}
which correlates with the conventional WKB approximation \cite{landau,
child} in the classically allowed region where $Q(r)>0$.  As is shown by
Fig.\,1, in contrast to the original wavefunction $\psi(r)$, the modified
wavefunction $\phi(y)$ has loops of almost constant amplitude and spacing
over the potential well, behavior which is qualitatively very similar to
that implied by Eq.\,(\ref{eq:sincos}).  It is therefore expected that
the WKB mapping of Eq.\,(\ref{eq:WKB}) should be close to ``optimal"
for all numerical methods (such as the finite-difference and collocation
methods) that are based on an equidistant grid in the classical region
of motion.  It should be stressed, however, that the mapping function
defined by Eq.\,(\ref{eq:WKB}) cannot itself be applied for scattering
length calculations because it does not satisfy the required asymptotic
behavior of Eq.\,(\ref{eq:ry_to_b}), and because it becomes imaginary in
the classically forbidden region where $r<r_0$.

It is easy seen from Eqs.\,(\ref{eq:gscatogilvie}) and (\ref{eq:arctag})
that the density function $\rho_{\texttt{OT}}(r)$ corresponding to the
one-parameter mapping function of Eq.\,(\ref{eq:ogilvie}) is proportional
to $1/[r+\bar{r}]^2$, while that for the two-parameters tangent function
of Eq.\,(\ref{eq:rarctan}) has a Lorentzian form with a maximum at
$\bar{r}$.  The plots presented in Fig.\,\ref{fig:Cs2dens},
show that there is good agreement between the $\rho_{\rm
WKB}(r)$ and $\rho_{\rm tg}(r)$ density functions over much of the domain of
the wavefunction $\psi(r)$. This leads to the conclusion that the optimal
value of parameter $\bar{r}_{\rm opt}$ should be close to the equilibrium
internuclear distance $r_e$ of the potential $U(r)$.  Hence, for the
two-parameters mapping function (\ref{eq:rarctan}), it seems reasonably to fix
$\bar{r}_{\rm opt}\approx r_e$ and to vary only the single parameter $\beta$.
Note that the pronounced differences among the density functions for
these three cases at small distances is not very important, because the
associated wave function $\psi(r)$ dies off exponentially in this region.

In the following, therefore, parameter $\bar{r}$ of the mapping
functions $y_{\texttt{OT}}(r;\bar{r})$ of Eq.\,(\ref{eq:ogilvie}) and
$y_{\rm tg}(r;\bar{r},\beta)$ of Eq.\,(\ref{eq:rarctan}) is fixed as
$\bar{r}=r_e$, while the optimum value of $\beta$ of Eq.\,(\ref{eq:rarctan})
is determined from a one-dimensional minimization of the functional
of Eq.\,(\ref{eq:minRichardson}).

%
%

\section{IMPLEMENTATION, TESTING, AND DISCUSSION}\label{numtest}

\subsection{Model-Potential Applications and Tests }

A computer program based on the adaptive mapping procedure described above
has been written and tested \cite{EPAPS}, both for a variety of model
potentials, and on potentials for real systems taken from the literature.
All results were obtained on 32 bit processors, mainly using double-precision
arithmetic, but with quadruple-precision calculations being used sometimes
in order to delineate the impact of accumulated numerical round-off error.
This section presents the results of those illustrative applications,
together with some general discussion of the method.

\begin{table}[t]
\caption{The $s$-wave scattering wavelengths $a_s$ (in \AA)
calculated for the LJ($2n,n$) potentials defined by Eq.(\ref{eq:ULJ}) \cite{EPAPS}.  All models have
equilibrium distance $r_e=1$\,[\AA] and the reduced mass is set
as $\mu=16.85762920$\,[au] so that in ``spectroscopists units", the
scaling factor $\displaystyle{\nicefrac{\hslash^2}{2\mu}}= 1$\, [${\textrm
{cm}}^{-1}\,{\textrm \AA}^2$]. ${\mathfrak D}_e$ is the well depth in
cm$^{-1}$, while $v_{\textrm {max}}$ is the vibrational quantum number of
the last bound level supported by the potential. \label{tab:LJ}}
\begin{ruledtabular}
\begin{tabular}{c c c c c c c} 
$n$ & $v_{\textrm {max}}$ & ${\mathfrak D}_e$ & $\beta_{\rm opt}$ & $\frac{n}{2}-1$
& $a_s^{\ddag}$ & $a_s$\\
\noalign{\smallskip}
\hline
\noalign{\smallskip}
4 & 14 & 1000 & 0.9 & 1.0 & 277.4 & ~~~310.54293138289\\
5 & 14 & 2165 & 1.5 & 1.5 & 234.9 & ~~~246.72686552846\\
6 & 14 & 3761 & 2.1 & 2.0 & 236.1 & ~~~242.48308194261\\
\noalign{\smallskip}
6 & 99 & 176200 & 2.1 & 2.0 & &~~~10.849479064634\\
6 & 99 & 174370 & 2.1 & 2.0 & &~~~11552.057690297\\
\end{tabular}
\end{ruledtabular}
$^{\ddag}$The $a_s$ estimates obtained using the last-level binding energy approximation of Eq.(\ref{eq:Energylast}).
\end{table}

The first sample calculations presented here are for the model
Lennard-Jones($2n,n$) potentials
\begin{eqnarray}\label{eq:ULJ}
U_{\textrm {LJ}}(r)~=~{\mathfrak D}_e \left[ \left(\frac{r_e}{r}
\right)^{2n} -~2 \left( \frac{r_e}{r} \right)^n \right] ~~.
\end{eqnarray}
defined by the potential function parameters and system reduced mass listed
in Table \ref{tab:LJ}.  The first three of these LJ($2n,n$) potentials, each
supporting 15 bound vibrational levels, are the same models systems considered
in our recent application of this adaptive mapping approach to bound-state
problems \cite{meshkov08}.  The last two LJ($12,6$) potentials have much
deeper wells and support 100 vibrational levels.  They were introduced here to
highlight the efficiency of the present method, since the computational
effort of scattering-length calculations increases dramatically as
$v_{\textrm {max}}$ and the absolute magnitude of $a_s$-values increase.

\begin{figure}
\includegraphics{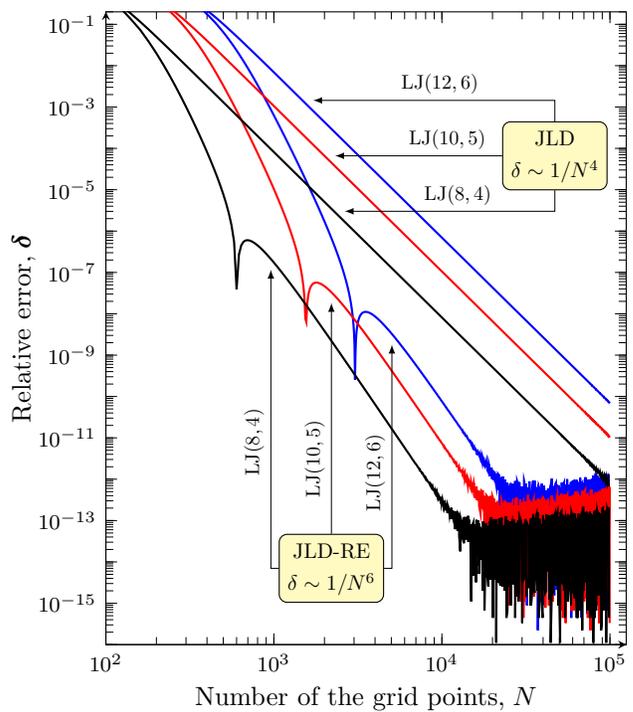}    \vspace{0mm}
\caption{(Color online) Convergence tests for $a_s$-values calculated in double precision arithmetic for the model 15-level LJ($2n,n$) potentials of Table~\ref{tab:LJ} using the one-parameter mapping function $y_{\texttt{OT}}(r;\bar{r})$ (\ref{eq:ogilvie}) with $\bar{r}=r_e=1$ [\AA]. Calculations performed using Johnson's log-derivative method alone are denoted `JLD', while the results labeled `JLD-RE' were obtained by also applying a ($N$,$N/2$)-Richardson extrapolation procedure to those results. \label{fig:LJOTconv}}
\end{figure}

\begin{figure}
\includegraphics{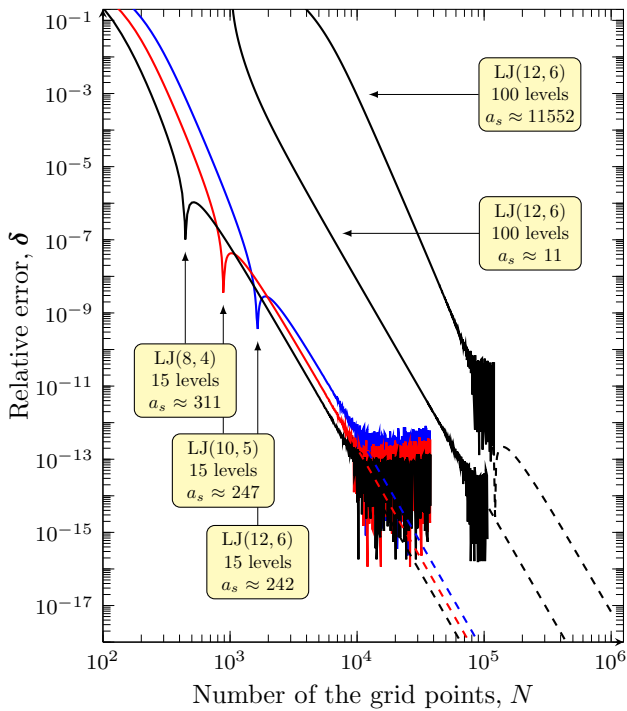}    \vspace{0mm}
\caption{(Color online) Convergence tests for $a_s$-values calculated by the JLD-RE($N$,$N/2$) method in double precision arithmetic (solid curves) for the five model LJ($2n,n$) potentials of Table~\ref{tab:LJ} using the two-parameters $y_{\rm tg}(r;\bar{r}=r_e,\beta)$ mapping function of Eq.\,(\ref{eq:rarctan}) with the optimal parameters $\beta_{\rm opt}$ listed there. The dashed lines present results obtained using quadruple-precision
arithmetic. \label{fig:LJ_conv}}
\end{figure}

Converged $a_s$-values for five model LJ($2n,n$) potentials are presented in
the last column of Table~\ref{tab:LJ}.  As discussed above, the range-mapping
parameter was fixed at $\bar{r}=r_e=1$[\AA] for both $y_{\texttt{OT}}(r)$
and $y_{\rm tg}(r)$ mapping functions.  The optimal $\beta_{\rm opt}$
values for the latter case were determined by minimizing the functional
of Eq.\,(\ref{eq:minRichardson}), yielding the results shown in the column
four. It is interesting to see that these empirically determined $\beta_{\rm
opt}$ values have essentially the same dependence on the inverse-power
$n$ governing the long-range behavior of the potential as was the case
for the analogous parameter $\alpha$ of the {\v S}urkus-variable mapping
(\ref{eq:surkus1}) used in the bound-state study of Ref.\,\cite{meshkov08}.


The log-log plots in Figs,\,\ref{fig:LJOTconv} and \ref{fig:LJ_conv} show
how the relative errors in the calculated $a_s$ values
\begin{equation} \label{eq:delta}
\delta = \left| \frac{a_s^{\rm calc}}{a_s^{\rm exact}} -1 \right|
\end{equation}
depend on the number of mesh points $N$ used in the radial integration.
The reference values $a_s^{\rm exact}$ were obtained from calculations using
quadruple precision arithmetic by increasing $N$ until the full desired
level of convergence was achieved.  The numerical `noise' on these plots
for $\delta \lesssim 10^{-12}$ indicates the precision limits achieved
using ordinary double-precision arithmetic, while the dashed lines show
how the convergence trend continues when quadruple-precision arithmetic
is used.

The three upper curves in Fig.\,\ref{fig:LJOTconv} (labeled JLD)
display results obtained using Johnson's log-derivative method, while the
three lower curves illustrate the greatly improved convergence achieved
when that approach is coupled to the ($N,N/2$) Richardson extrapolation
procedure described in the Appendix.  These results clearly confirm the
prediction of Eq.\,(\ref{eq:Richardson}), that Johnson's log-derivative
method (JLD), and that method combined with a Richardson extrapolation
to zero step (JLD-RE), demonstrate $N^{-4}$ and $N^{-6}$ convergence
rates, respectively.  We see that for these 15-level LJ potentials,
the JLD-RE($N,N/2$) method allows us to attain 11-12 significant digits
in calculated $a_s$ values when using only $N\approx 10^4$ radial mesh
points.

The results displayed in Fig.\,\ref{fig:LJ_conv} show that as expected, the
computational effort increases significantly for the 100-level LJ($2n,n$)
models, especially for the very last case considered in Table 1, for which
the scattering length is extremely large: $a_s\approx 10^4$\,[\AA].  However,
that even for those cases full double-precision converge is achieved with
only $N\approx 10^5$.  Note that for a given number of grid points, use
of the simple one-parameter mapping function $y_{\texttt{OT}}(r)$ yields
relative errors that are only $\sim 10$ times larger than the analogous
results yielded by the optimum two-parameter functions $y_{\rm tg}(r)$.
The non-linear `cusp'-like behavior near $N\approx 10^3$ on the JDL-RE
plots in Figs.\,\ref{fig:LJOTconv} and \ref{fig:LJ_conv} appears to be due
to accidental cancelation of higher-order terms in the RE series expansion
for these cases.  It does not appear in analogous results based on use of
a Numerov propagator.

\begin{figure}
\includegraphics{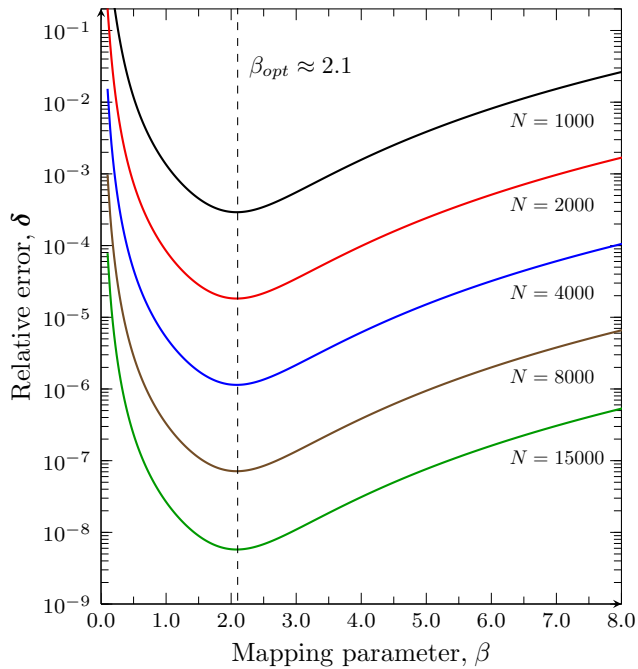}\vspace{0mm}
\caption{(Color online) Relative errors (\ref{eq:delta}) in the $a_s$-values
obtained for the 15-level LJ$(12,6)$ potential (\ref{eq:ULJ}) as functions of the mapping
parameter $\beta$, where $N$ is the number of grid points used in the JLD
method. \label{fig:optbeta}}
\end{figure}


The results presented in Fig.\,\ref{fig:optbeta} clearly show that the
accuracies of $a_s$ values calculated with any given number of mesh
points $N$ vary smoothly with the value of mapping parameter parameter
$\beta$.  They also show that the optimal parameter value $\beta_{\rm
opt}$ is essentially independent of the number of integration points used.
Analogous results are obtained when modeling LJ($2n,n$) potentials
with $n=4$ and 5.

\begin{figure}
\includegraphics{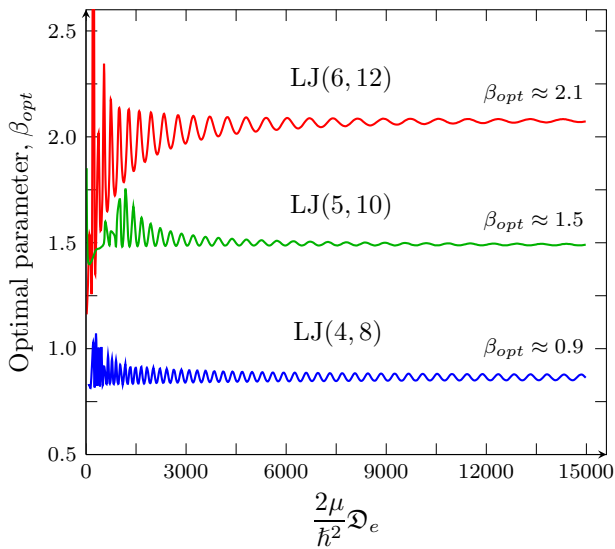}    \vspace{0mm}
\caption{(Color online) Optimal mapping parameter $\beta_{\rm opt}$ as a
function of the dissociation energy ${\mathfrak D}_e$ of 15-level LJ($2n,n$)
potentials (\ref{eq:ULJ}) for $n=4$, 5 and 6.
\label{fig:betaDe}}
\end{figure}

The sensitivity of the value of $\beta_{\rm opt}$ to the depth of the
potential energy well has also been investigated.  Figure~\ref{fig:betaDe}
shows that the values of $\beta_{\rm opt}$ for model LJ($2n,n$) potentials
only depend significantly on the dissociation energy ${\mathfrak D}_e$ for
very shallow potentials which support only a few bound states.  For such
species (e.g., see the He$_2$ example considered below), accurate scattering
lengths can readily be calculated using a relatively small number of grid
points, so determining precise values of $\beta_{\rm opt}$ is immaterial.
Overall, Fig.\,\ref{fig:betaDe} shows that the optimal value of the $\beta$
parameter in the mapping function of Eq.\,(\ref{eq:rarctan}) depends mainly
on the inverse-power $n$ governing the limiting long-range behavior of
the interaction potential.  As indicated by Fig.~\ref{fig:betaDe} and
Table \ref{tab:LJ}, these optimal values are approximately defined by the
simple relationship
\begin{equation}  \label{eq:betaopt}
\beta_{\rm opt} ~\approx~ \frac{n}{2} - 1  ~~,
\end{equation}
at least for these LJ$(2n,n)$-like potentials.  We believe that this
empirical rule will be valid when applying the two-parameter mapping function
of Eq.\,(\ref{eq:rtan}) to any deeply bound interatomic potential having
a long-range tail which dies off as $1/r^{n}$ for $n\geq 4$.

\begin{figure}
\includegraphics{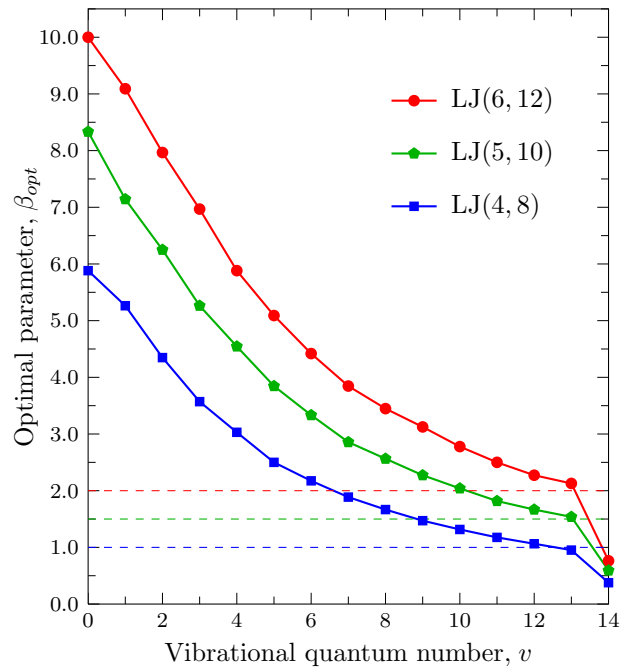}    \vspace{0mm}
\caption{(Color online) Optimal mapping parameters $\beta_{\rm opt}$
determined for the bound vibrational levels $v\in [0,14]$ of the
model 15-level LJ($2n,n$) n=4,5,6 potentials (\ref{eq:ULJ}). The
dashed horizontal lines corresponds to the $\beta_{\rm opt}$ values implied by
Eq.(\ref{eq:betaopt}).
\label{fig:betabound}}
\end{figure}

It is interesting to note that the two-parameters mapping function $y_{\rm
tg}(r;\bar{r},\beta)$ of Eq.\,(\ref{eq:rarctan}) is also well suited for
solving the Schr{\"o}dinger equation for the bound-state problem discussed
in Ref.\,\cite{meshkov08}.  In particular, Fig.\,\ref{fig:betabound} shows
how the values of $\beta_{\rm opt}$ evolve if this mapping parameter is
optimized independently for each level of our three 15-level LJ($2n,n$)
potentials.  Except for the very last level, the values of $\beta_{\rm opt}$
smoothly approach the limiting value implied by Eq.\,(\ref{eq:betaopt}) as
the vibrational levels approach dissociation.  The substantial deviation
from this limiting behavior for the lower vibrational levels is not a
matter of concern, since those levels can readily be located accurately
using a very wide range of $\beta$ values, and since their radial amplitude
is much smaller than that for the highest levels, the overall computational
efficiency for such levels is not very strongly dependent on the value of
$\beta$.  The abrupt drop-off in the values of $\beta_{\rm opt}$ for the
very last level also mimics the behavior found in Ref.\,\cite{meshkov08}
using very different mapping functions (\ref{eq:surkus2}) based on the
{\v S}urkus variable \cite{Surkus} defined by Eq.(\ref{eq:surkus1}).
This abrupt decrease of $\beta_{\rm opt}$ for the last very weakly
bound vibrational level $v=v_{\rm max}=14$ is related to the form of the
corresponding wavefunction $\psi_{v_{\rm max}}(r)$ which has very broad
last loop centered at a very large internuclear distance.  In particular,
for the case of a last level which lies extremely close to dissociation,
it appears that a small limiting value of $\beta_{\rm opt}\lesssim 0.5$
is needed to provide a sufficiently broad distribution of grid points to
properly characterize the outermost loop of the wavefunction.

\subsection{Applications to `Real' Systems}

This section describes application of our new method of performing scattering
length calculations to three `real' physical problems.  The first of these
is the elastic scattering of a free electron from a neutral Xe atom.
The function used for the Xe--e$^-$ interaction potential is the
`HFD'-type potential reported by Szmytkowski \cite{szmytkowski95} (with
energy and distance in atomic units):
\begin{eqnarray}\label{eq:HFDe}
U_{\textrm {Xe-e}^-}(r)~=~A\, e^{-\gamma r^2}-\sum_{n=4,6}F_n(r)\,
C_n/r^n ~~,
\end{eqnarray}
in which $A=306.0$ and $\gamma=1.0$, while the $n$-dependent damping
function $F_n$ has the form
\begin{eqnarray}\label{eq:dampC4C6}
F_n(r)=\left[1-e^{-(r/r_c)^2}\right]^n
\end{eqnarray}
with $r_c=1.89$ being a cut-off radius, such that $F_n(r)=0$ for
$r < r_c$.  Following Czuchaj {\em et al.}\ \cite{czuchaj87}, its
dispersion coefficients $C_4=\alpha_1/2$ and $C_6=(\alpha_2-6\beta_1)/2$
in Eq.(\ref{eq:HFDe}) correspond to the familiar charge/induced-dipole
and charge/induced-quadrupole interactions, where $\alpha_1=27.292$ and
$\alpha_2=128.255$ are the static dipole and quadrupole polarizabilities of
the Xe atom, while $\beta_1=29.2$ is the dynamical correction to the dipole
polarizability.  This shallow $U_{\textrm{Xe-e}^-}(r)$ interaction potential
supports no bound levels and its scattering length is negative, as was
determined by Szmytkowski using an asymptotic method \cite{szmytkowski95}.

The present $a_s$-value for this system (in $au$), given in the first row
of Table \ref{tab:scatlen}, was converged to 14 significant digits using
the JLD-RE($N$,$N/2$) method based on $N\approx 10^3$ grid points and a scaling factor
of $\hslash^2/2\mu=0.5$.  However, only $N\approx 100$ grid points are
required to converge the present $a_s$-value to 7 significant digits. The
present estimate agrees with the value reported by Szmytkowski
\cite{szmytkowski95} to within the 6 significant digit that he reports
(see row 2 of Table \ref{tab:scatlen}).

\begin{table}[t]
\caption{Comparison of $s$-wave scattering lengths $a_s$ (in $au$) obtained in the framework of JLD-RE($N$,$N/2$) 
procedure \cite{EPAPS} using realistic published potentials for Xe--e$^-$ \cite{czuchaj87}, Cs$_2(a\,^3\Sigma_u^+)$ 
\cite{gribakin93, marinescu94, szmytkowski95}, and $^{3,4}$He$_2(X^1\Sigma_g^+)$ \cite{aziz79} 
interactions and two-parameter mapping function $y_{\rm tg}(r;\bar{r},\beta)$ of Eq.\,(\ref{eq:rarctan}). 
The optimal mapping parameters $\beta_{\rm opt}$ were determined by minimization the functional
(\ref{eq:minRichardson}), while the $\bar{r}$ values (in $au$) were fixed
at $\bar{r}_{\rm opt}=r_e$, where $r_e$ is the equilibrium distance of
the relevant potential.  \label{tab:scatlen}}
\begin{ruledtabular}
\begin{tabular}{cccclcc}
species & potential & n & $v_{\rm max}$ & \multicolumn{1}{c}{$a_s$} & $\beta_{\rm opt}$ & $\bar{r}_{\rm opt}$\\
\noalign{\smallskip}
\hline
\noalign{\smallskip}
Xe--e$^{-}$ & Eq.\,(\ref{eq:HFDe}) & 4 & $^{\ddag}$ & ~~-4.95280521509712 & 1.7& 3.2\\
            &                      &   &   & ~~-4.95281\footnote{Ref.\,\cite{szmytkowski95}} \\

\noalign{\smallskip}
Cs$_2$& Eq.\,(\ref{eq:Cs2})& 6 & 57 & ~~68.215967213 & 2.0 & 12.0\\
                         & &   &    & ~~68.21596\footnote{Ref.\,\cite{marinescu94}} \\
                         & &   &    & ~~68.21823$^a$       \\
                         & &   &    & ~~68.0\footnote{Ref.\,\cite{gribakin93}}  \\
\noalign{\smallskip}
$^{4}$He$_2$& Eq.\,(\ref{eq:He2})& 6 & 0 & 236.3688418603     & 2.5 & 5.67\\
            &                    &   &   & 236.36884174\footnote{Ref.\,\cite{gutierrez84}}\\
\noalign{\smallskip}
$^{3}$He$_2$&                    &   & $^{\ddag}$ & -13.17845206095 & 3.6 & 5.67\\
            &                    &   &   & -13.178452062$^{d}$  \\

\end{tabular}
\end{ruledtabular}
$^{\ddag}$There are not bound levels.
\end{table}

Our second real-world example is a modified 'Hartree-Fock
dispersion' (HFD) type potential for the $a\,^3\Sigma_u^+$ ground triplet
state of the cesium dimer \cite{gribakin93, marinescu94, szmytkowski95}
(again, with energies and distances in atomic units):
\begin{eqnarray}\label{eq:Cs2}
U_{{\textrm Cs}_2}(r)~=~A\,r^{\alpha}\, e^{-\gamma r}-F(r)\left[
   \sum_{n=6,8,10}\! C_n/r^n \right] ~~.~~~~~
\end{eqnarray}
The first term in Eq.(\ref{eq:Cs2}), defined by constants $A=8\times
10^{-4}$, $\alpha=5.53$ and $\gamma=1.072$, represents the exchange repulsion
energy, while the second is a sum of van der Waals dispersion terms (with
coefficients $C_6=7.02\times 10^3$, $C_8=1.1\times 10^6$, $C_{10}=1.7\times
10^8$) multiplied by the $n$-independent damping function
\begin{eqnarray}  \label{eq:dampC6C8C10}
F(r)~=~H(r-r_c)+H(r_c-r)\,e^{-(r_c/r-1)^2}
\end{eqnarray}
where $H(x)$ is the Heaviside step function: $H(x)=1(0)$, when $x\geq(<)0$,
while $r_c=23.165$ is the cut-off radius. This potential energy function
supports up to $58$ bound levels, and can be considered a typical
example of a many-level interatomic potential.

The present $a_s$-value, obtained using our JLD-RE($N$,$N/2$) procedure with
a scaling factor of $\hslash^2/2\mu=1/2.422\times 10^5$, is listed in row 3
of Table \ref{tab:scatlen}. It agrees with previous estimates obtained
using WKB \cite{gribakin93} (row 6), asymptotic \cite{szmytkowski95}
(row 5), and iterative \cite{marinescu94} (row 4) methods to within 2,
4 and 7 significant digits, respectively. The present method required
about $N\approx 10^4$ grid points to obtain 10 significant digits in the
calculated $a_s$-value.

Our third practical application is to the ground $X\,^1\Sigma_g^+$ state
of the helium dimer, for which the interaction potential
is again represented by an `HFD'-type function \cite{aziz79} written as (with
energies in K and distances in \AA)
\begin{eqnarray}\label{eq:He2}
U_{{\textrm He}_2}(r)={\mathfrak D}\left[ A\exp(-\gamma x)-F(x)\left(\sum_n
\frac{C_n}{x^n} \right)\right]~~,~~~
\end{eqnarray}
in which $x=r/r_m$,  ${\mathfrak D}=10.8$, $A=544850.4$, $\gamma=13.353384$,
$C_6=1.3732412$, $C_8=0.4253785$, $C_{10}=0.1781$, and $r_m=2.9673$.
The damping function $F(x)$ in Eq.\,(\ref{eq:He2}) is defined by
Eq.(\ref{eq:dampC6C8C10}) with the parameter $x_c=r_c/r_m =1.28$.  This is
an exotic example of very shallow interatomic potential, as it supports only
a single bound level for the heavier isotopologue $^4$He$_2$ and no bound
levels at all for the lighter isotopologue $^3$He$_2$.  For consistency with
the calculations of Ref.\,\cite{gutierrez84}, the values of the scaling
factors $\hslash^2/2\mu$ used in the present calculation were $16.085775$
and $12.120904$ for $^3$He$_2$ and $^4$He$_2$, respectively.

The results presented in the last four rows of Table \ref{tab:scatlen}
show that the present $a_s$-values (first entry for each case) agree with
the result obtained in Ref.\cite{gutierrez84} using the asymptotic method
to about 10 significant digits.  However, the JLD-RE($N,N/2$) procedure
used here required only $N=100$ grid points to obtain $a_s$-values for
both isotopologues with uncertainties of only 0.01\%.

\begin{figure}[t!]   
\includegraphics{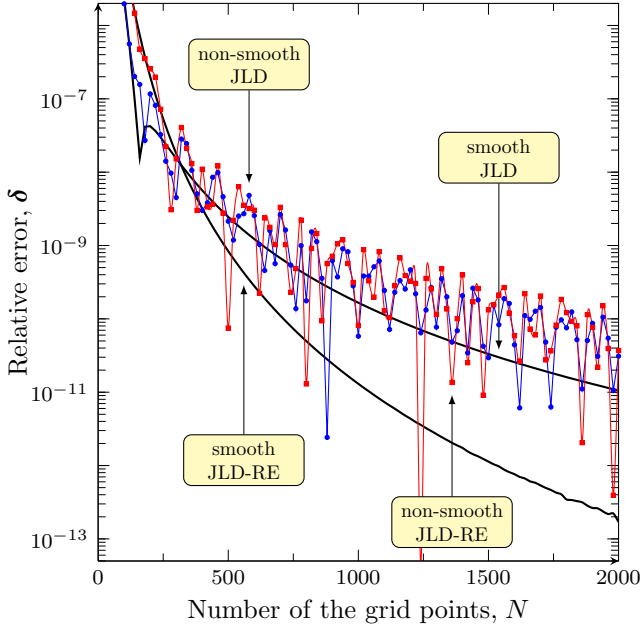}    \vspace{0mm}
\caption{(Color online) Comparison of convergence behavior for scattering
lengths implied by the $^3$He$_2(X\,^1\Sigma_g^+)$ potential \cite{aziz79}
as calculated by the JLD($N$) and JLD-RE($N,N/2$) procedures using smooth {\em vs}\
non-smooth long-range damping functions of Eq.\,(\ref{eq:dampC6C8C10}).}
\label{fig:He2conv}
\end{figure}

An interesting general point concerns the fact that the convergence
behavior of the $a_s$-values calculated for Cs$_2$ and He$_2$ at
high $N$ was not as smooth as it was for the model LJ($2n,n$) and
realistic Xe--e$^-$ potentials; this is demonstrated by the contrast
between the curves in Fig.\,\ref{fig:LJ_conv} and the 'non-smooth' 
curves in Fig.\,\ref{fig:He2conv}.  We attribute this behavior to the
discontinuous second and higher derivatives of the damping function
$F(x)$ in the potentials of Eqs.\,(\ref{eq:Cs2}) and (\ref{eq:He2}), at
the point $r=r_c$.  To confirm this assertion, the $a_s$ calculations
for $^3$He$_2$ were repeated with the Heaviside switching function of
Eq.\,(\ref{eq:dampC6C8C10}) replaced by the smooth switching function
\begin{eqnarray}\label{eq:Heavisideasymptot}
\tilde{H}(x)=\frac{1}{1+e^{-2kx}};\qquad k=10
\end{eqnarray}
The results obtained in this way, plotted as the 'smooth' curves in
Fig.\,\ref{fig:He2conv}, clearly confirm our assertion.  This demonstrates
the importance of having potential energy functions models which have a
very high degree of analytic smoothness.

\begin{figure}[t!]   
\includegraphics{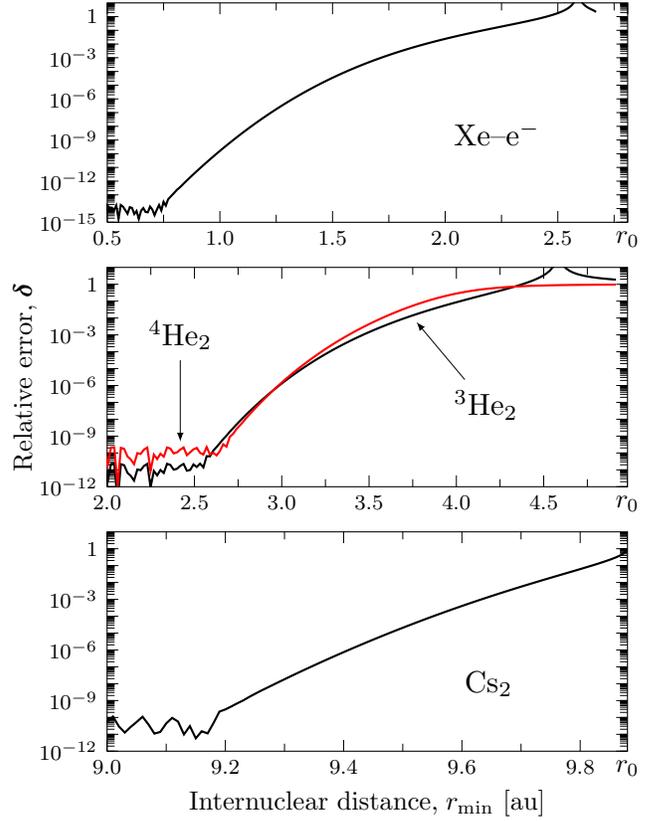}   \vspace{-2mm}
\caption{(Color online) Convergence of calculated scattering lengths for
the Xe--e$^{\textrm-}$, $^{3,4}$He$_2(X\,^1\Sigma_g^+)$, and
Cs$_2(a\,^3\Sigma_u^+)$ systems with respect to the lower bound of the
integration interval, $r_{\rm min}$, where $r_0$ is the distance where
$U(r)=0$.}\label{fig:Rmin}
\end{figure}

Our final point concerns the sensitivity of the calculated $a_s$-values
to the position of the inner boundary $r_{\rm min}(y=a)$ of the
integration region.  While the transformation of Eqs.\,(\ref{eq:rtan}) and
(\ref{eq:rarctan}) formally fixes the lower bound for $y_{\rm tg}(r)$ to
correspond to $r=0$, the exponentially rapid decay of the wavefunction in
the classically forbidden region under the short-range repulsive potential
wall means that in practice that lower bound may be set quite a bit closer
to the left turning point $r_0$ where $U(r)=0$.  To examine this point,
Fig.\,\ref{fig:Rmin} shows the convergence behavior of calculated values
of $a_s$ for the three real systems of Table \ref{tab:scatlen} as the lower
limit of the integration interval is shifted inward from the distance $r_0$.
It was found that values of $r_{\rm min}$ for which the relative error 
approaches the double-precision numerical-noise limit can be
defined by the criterion
\begin{eqnarray}\label{eq:rmin}
(r_{\rm min}-r_0)\times \sqrt{\frac{2\mu}{\hbar^2}|U(r_{\rm min})|}~
\gtrsim ~23 ~~,
\end{eqnarray}
which is based on semi-classical exponential decay of the original
$\psi(r)$ wavefunction in the classically forbidden region by a factor
of $\sim 10^{-10}$.  For the $y_{\rm tg}(r;\bar{r},\beta)$ mapping function of
Eq.(\ref{eq:rarctan}), this implies an inner (left hand) boundary
point of $a=\frac{2}{\pi}~\tan^{-1}\left[\beta\left(r_{\rm
min}/\bar{r}-1\right)\right]\,$.

\section{CONCLUSIONS}

This paper presents a very robust and efficient new way of
calculating scattering lengths.  The two-parameter mapping
function of Eq.\,(\ref{eq:rtan}) transforms the conventional radial
Schr{\"o}dinger equation (\ref{eq:schrod0}) into the equivalent form
of Eq.\,(\ref{eq:schrod}) defined on the finite domain $y\in [a,1]$.
For arbitrary interaction potentials, neither the solution $\phi(y)$ of
the transformed equation nor its first derivative $\phi'(y)$ are singular
anywhere on this interval.  As a result, the $s$-wave scattering length
$a_s$ can be exactly expressed in terms of the logarithmic derivative of
this transformed wave function $\phi(y=1)$ at the right boundary point,
as specified by Eq.\,(\ref{eq:scatarctan}). The method does not depend on
a particular asymptotic form of the potential as long as it dies off as
fast as $1/r^{n}$ for $n\geq 4$, as $r\to \infty$.

For well-bound potentials with equilibrium distance $r_e$ and a limiting
(attractive) long-range behavior of $1/r^n$, the optimal values of the
$y_{\texttt{tg}}(r)$ mapping parameters have been shown to be $\bar{r}\approx
r_e$ and $\beta\approx \frac{n}{2}-1$, respectively.  These same mapping
parameters also yield efficient solutions when this approach is used for
solving bound-state problems.  Regardless of the absolute magnitude of the
scattering length or the number of bound levels supported by the potential,
the accuracy of the $a_s$ calculation can easily achieve 10-12 significant
digits when working in ordinary double-precision arithmetic. It is also
shown that stable and highly precise values of $a_s$ cannot be determined for
analytical potentials which lack high-order analytic smoothness throughout
the classically allowed region.  A computer program applying this approach has
been submitted to the Journal's online data archive \cite{EPAPS}.

Finally, we note that although it requires a little more computational
effort to achieve a given level of precision, combination of the
$y_{\texttt{OT}}(r)$ mapping function of Eq.\,(\ref{eq:ogilvie}) with
a conventional Numerov wavefunction propagator \cite{Numerov} can yield
both accurate $a_s$ values and wavefunctions $\phi(y)$ that can be used
for calculating photoassociation cross sections and other properties
\cite{LEVELas}.

\begin{acknowledgments}

This work has been supported by the Russian Foundation for Basic Research
by grant 10-03-00195a, and by NSERC Canada. The Moscow team is also
grateful for partial support from the Federal Program "Scientists and
Educators for an Innovative Russia 2009-2013", contract P 2280.

\end{acknowledgments}


\appendix*

\section{Johnson's log-derivative method}

As was shown in Refs.\,\cite{johnson73} and \cite{johnson77}, Johnson's
quadrature procedure for outward integration of the Riccati equation
(\ref{eq:riccati}) is based on the two-point finite-difference scheme:
\begin{eqnarray} \label{eq:johnson}
z_k~=~\frac{z_{k-1}}{1+z_{k-1}}~-~\left (\frac{h^2}{3}\right)w_{k}u_{k} ~~,
\end{eqnarray}
where $\xi_k(y_k)=h^{-1}z_k(y_k)$, the mesh points are $y_k=a+kh$, the
integration step is $h=(b-a)/N$, and
\begin{eqnarray} \label{eq:johnson1}
u_k= \left\{
\begin{array}{ll}
\widetilde{Q}(y_k)& k=0,2,4,\ \ldots\,,N \\[1ex]
\frac{\widetilde{Q}(y_k)}{1+(h^2/6)\widetilde{Q}(y_k)}~ & k=1,3,5,\,\ldots\,,N-1~~
\end{array}
\right.
\end{eqnarray}
with weights $w_k$ as in a Simpson quadrature
\begin{eqnarray} \label{eq:johnson2}
w_k~=~ \left\{
\begin{array}{ll}
1\,~& k=0,N \\[1ex]
4& k=1,3,5,\,\ldots\, ,N-1\\[1ex]
2& k=2,4,6,\,\ldots\, ,N-2
\end{array}
\right.
\end{eqnarray}
The total number of integration points must be odd, so $N$ must be an
even number.

In the classically forbidden region where $\widetilde{Q}(y_0)<0$, the
initial value log-derivative solution $z_0=z(a)$ can be estimated using the
semi-classical approximation \cite{child, johnson77}:
\begin{eqnarray} \label{eq:johnsonWKB}
z_0 = h\left[\sqrt{-\widetilde{Q}(y_0)}-\left(\frac{h}{3}\right )\widetilde{Q}(y_0)\right]
\end{eqnarray}

It has been verified by numerical calculations \cite{johnson77} that the
cumulative truncation error of the log-derivative method is given by
\begin{eqnarray} \label{eq:Richardson}
\xi_{h\to 0}-\xi_{h}~=~C\,h^4+{\mathcal O}(h^6)
\end{eqnarray}
As a result, an approximate solution $\xi_h(b)$ determined with fixed
integration stepsize (or a fixed number of mesh points) $h_i(N_i)$ can be
extrapolated to zero step size using Richardson's formula \cite{Richardson,
NR}:
\begin{equation} \label{eq:Richardson1}
\xi_{h\to 0}(b)~\simeq~ \xi_{h_1}(b)+\Delta \xi(b) ~~,
\end{equation}
where
\begin{equation} \label{eq:Richardson2}
\Delta \xi~=~\frac{\xi_{h_1}-\xi_{h_2}}{\lambda^4-1}~~;\quad
\lambda~\equiv~\frac{h_2}{h_1}~=~\frac{N_1}{N_2} ~~.
\end{equation}


\begin{thebibliography}{99}
\bibitem{landau} L. D. Landau and E. M. Lifshitz, \emph{Quantum Mechanics:
Non-relativistic theory}, Pergamon Press (1965).

\bibitem{mott} N. F. Mott and H. S. W. Massey, \emph{The Theory of Atomic
Collisions}, Oxford University Press (1965).

\bibitem{bloch08} I. Bloch, J. Dalibard, and W. Zwerger, Rev. Mod. Phys.,
\textbf{80}, 885 (2008).

\bibitem{hutson06} J.M. Hutson and P.Soldan, Int. Rev. Phys. Chem.,
\textbf{25}, 497 (2006).

\bibitem{pethick02} C. Pethick and H. Smith, \emph{Bose-Einstein Condensation
in Dilute Gases}, Cambridge University Press (2002).

\bibitem{rodberg67} L. S. Rodberg and R. M. Thaler, \emph{Introduction to
the Quantum Theory of Scattering}, Academic Press, New York (1967).

\bibitem{roman65} P. Roman, \emph{Advanced Quantum Theory: An Outline of
the Fundamental Ideas}, Addison-Wesley, Reading, Massachusetts (1965).

\bibitem{jones06} K. M. Jones, and E. Tiesinga, P.D. Lett, and
P. S. Julienne, Rev. Mod. Phys., \textbf{78}, 483 (2006).

\bibitem{Abraham95} E. R. I. Abraham,  W. I. McAlexander, C. A. Sackett,
and R. G. Hullet, Phys. Rev. Lett., \textbf{74}, 1315 (1995).


\bibitem{Julienne1995} O.Dulieu and P.S. Julienne, J. Chem. Phys., \textbf{103}, 60 (1995).

\bibitem{gutierrez84} G. Guti\'errez,  M. de Llano, and W. C. Stwalley,
Phys. Rev. B, \textbf{29}, 5211 (1984).

\bibitem{marinescu94} M. Marinescu, Phys. Rev. A, \textbf{50}, 3177 
(1994).

\bibitem{szmytkowski95} R. Szmytkowski, J. Phys. A: Math. Gen., \textbf{28},
7333 (1995).

\bibitem{ouerdane03} H. Ouerdane, M. J. Jamieson, D. Vrinceanu and
M. J. Cavagnero, J. Phys. B: At. Mol. Opt. Phys., \textbf{36}, 4055 (2003).

\bibitem{gribakin93} G. F. Gribakin and V. V. Flambaum, Phys.\ Rev.\ A, \textbf{48}, 546 (1993).

\bibitem{child} M.S. Child, {\em Semiclassical Mechanics with Molecular
Applications}, Clarendon Press, Oxford (1991).

\bibitem{LiouvilleGreen} J.\ Liouville, J.\ Math.\ Pure Appl., {\bf 2}, 16
(1837); G.\ Green, Trans.\ Cambridge Phil.\ Soc., {\bf 6}, 457 (1837).

\bibitem{meshkov08} V. V. Meshkov, A. V. Stolyarov, and R. J. Le Roy,
Phys. Rev. A, \textbf{78}, 052510 (2008).

\bibitem{Ogilvie} J. F. Ogilvie, Proc. Roy. Soc. (London) A, \textbf{378},
287 (1981).

\bibitem{Surkus} A.\ A.\ Surkus, R.\ J.\ Rakauskas, and A.\ B.\ Bolotin, 
Chem.\ Phys.\ Lett., \textbf{105}, 291 (1984).

\bibitem{Numerov} B. Numerov, Publs. Obs. Cent. Astrophys. Russ., \textbf{2},
188 (1933).

\bibitem{pseudospectral} J. P. Boyd, \emph{Chebyshev and Fourier Spectral
Methods}, DOVER Publications, Inc., New York (2000).

\bibitem{johnson73} B.R. Johnson, J. Comp. Phys., \textbf{13}, 445 (1973).

\bibitem{johnson77} B.R. Johnson, J. Chem. Phys., \textbf{67}, 4086 (1977).

\bibitem{Richardson} L. F. Richardson, Phil. Trans. Roy. Soc. (London) A, 
\textbf{226}, 299 (1927).

\bibitem{NR} W.\ H.\ Press, S.\ A.\ Teukolsky, W.\ T.\ Vetterling, and B.\
P.\ Flannery,\emph{Numerical Recipes in Fortran 77}, Cambridge University
Press (1999).

\bibitem{czuchaj87} E. Czuchaj, J.Sienkiewicz and W. Miklaszewski,
Chem. Phys., \textbf{116}, 69 (1987).

\bibitem{aziz79} R. A. Aziz, V. P. S. Nain, J. S. Carley, W. L. Taylor
and G. T. McConville, J. Chem. Phys., \textbf{70}, 4330 (1979).

\bibitem{EPAPS} A FORTRAN 77 code for performing the $s$-wave scattering length
calculation based on the adaptive mapping procedure
has been deposited in the journal's on-line electronic
archive.  E-PAPS document files can be retrieved via the EPAPS homepage
(http://www.aip.org/epaps/epaps.html) or from ftp.aip.org in the
directory/epaps/.  See the EPAPS homepage for more information.

\bibitem{LEVELas} R. J. Le Roy, {\em  A Computer Program for Solving the
Radial Schr{\" o}dinger Equation for Bound and Quasibound Levels and
Scattering Lengths}, University of Waterloo Chemical Physics Research
Report CP-XXX (2011);  see \texttt{http://leroy.uwaterloo.ca/programs/}.

\end{thebibliography}

\end{document}